\documentclass{amsart}
\usepackage{amsfonts}
\usepackage{amsmath}
\usepackage{amssymb}
\usepackage{amsthm}
\usepackage{color}
\usepackage{comment}
\usepackage[foot]{amsaddr}

 \usepackage[utf8]{inputenc} 
\usepackage{pdfpages}

\usepackage[numbers]{natbib}

\definecolor{Green}{rgb}{0.20,0.43,0.09}

\newtheorem{Thm}{Theorem}

\theoremstyle{definition}

\theoremstyle{remark}

\numberwithin{equation}{section}

\def\bigO{\mathcal{O}}

\title[How to share a cake with a secret agent]{How to share a cake with a secret agent}
\date{\today}

\author[G.~Ch\`eze]{Guillaume Ch\`eze}
\address{Guillaume Ch\`eze: Institut de Math\'ematiques de Toulouse, UMR 5219\\
Université de Toulouse ; CNRS  \\
UPS IMT, F-31062 Toulouse Cedex 9, France 
}
\email{guillaume.cheze@math.univ-toulouse.fr}

\date{\today}

\begin{document}
	
\begin{abstract}

In this note we study a problem of fair division in the absence of full information. We give an algorithm which solves the following problem: $n\geq 2$ persons want to cut a cake into $n$ shares so that each person will get at least $1/n$ of the cake for his or her own measure, furthermore the preferences of one person are secret. How can we construct such shares?\\
Our algorithm is a slight modification of the Even-Paz algorithm and allows to give a connected part to each agent. Moreover, the number of cuts used during the algorithm is optimal: $\bigO\big(n\log(n)\big)$.\\
\end{abstract}

\maketitle

\section*{Introduction}
Fair division is an old problem. The following situation already appears in the Bible. Two persons (in the Bible Abraham and Lot) want to share a land. Furthermore, this division must be fair. This means that each agent think that he or she has obtained at least $1/2$ of the land for his or her own point of view. The following protocol, called ``Cut and Choose", is then used:\\
The first player cuts the land into two pieces with equal values for him or her. The second player chooses one of the two pieces.\\
With this strategy each player get a connected piece with a value bigger than $1/2$ for his or her point of view.\\
This protocol can also be  used to divide a cake or an heterogeneous good as time or computer memory between two agents.
How can we generalize this protocol to $n$ agents? \\
Several answers exist. In order to describe them we need to precise some points.\\
 
We consider an heterogeneous good, for example: a cake, represented by the interval $X=[0,1]$ and $n$ agents with different points of view. We associate to each agent a non-atomic probability measure $\mu_i$ on the interval $X=[0;1]$. These measures represent the utility functions of the agent. This means that if $[a,b] \subset X$ is a part of the cake then $\mu_i([a,b])$ is the value associated by the $i$-th agent to this part of the cake. As $\mu_i$ are probability measures, we have $\mu_i(X)=1$ for all  $i$.\\
A division of $X$ is a partition $X=\sqcup_j X_j$ where each $X_j$ is given to one agent. Thus there exists a permutation $\sigma \in \mathcal{S}_n$ such that $\mu_i$ is associated to $X_{\sigma(i)}$. A division is \emph{simple} when each $X_i$ is an interval.\\

 Several notions of fair division exists.\\
We say that a division is \emph{proportional} when $\mu_i(X_{\sigma(i)}) \geq 1/n$.\\
We say that a division is \emph{envy-free} when for $i\neq j$, we have $\mu_i(X_{\sigma(i)}) \geq \mu_i(X_{\sigma(j)})$.\\

 The problem of fair division (theoretical existence of fair division and construction of algorithms) has been studied in several papers \cite{Steinhaus,DubinsSpanier, EvenPaz, EdmondsPruhs, BramsTaylorarticle, RoberstonWebbarticle, AzizMackenzie}, and books about this topic, see e.g. \cite{RobertsonWebb,BramsTaylor, Procacciachapter,Barbanel}. These results appear in the mathematics, economics, political science, artificial intelligence and computer science literature. Recently, the cake cutting problem has been studied intensively by computer scientists for solving resource allocation problems in multi agents systems, see e.g.~\cite{Chevaleyre06}. \\
 
 In this note we are going to study proportional fair divisions. This topic has been very studied. Several algorithms already exist, see e.g. \cite{RobertsonWebb}.
  In order to describe algorithms we thus need a model of computation. There exist two main classes of cake cutting algorithms: discrete and continuous protocols (also called moving knife methods). Here, we study only discrete algorithms. These kinds of algorithms can be  described thanks to the  classical model introduced by Robertson and Webb and formalized by Woeginger and Sgall in \cite{Woeg}. In this model we suppose that a mediator interacts with the agents. The mediator asks two type of queries: either cutting a piece with a given value, or evaluating a given piece. More precisely, the two type of queries allowed are:
\begin{enumerate}
\item $eval_i(x,y)$: Ask agent $i$ to evaluate the interval $[x,y]$. This means compute $\mu_i([x,y])$.
\item $cut_i(x,\alpha)$: Asks agent $i$ to cut a piece of cake $[x,y]$ such that $\mu_i([x,y])=\alpha$. This means: for given $x$ and $\alpha$, solve $\mu_i([x,y])=\alpha$.
\end{enumerate} 
We remark that in the ``Cut and Choose" algorithm only these two kinds of queries are used.\\
In the Robertson-Webb model the mediator can adapt the queries from the previous answers given by the players. In this model, the complexity counts the finite number of queries necessary to get a fair division. For a rigorous description of this model we can consult: \cite{Woeg}\\

An optimal algorithm for proportional fair division has been given by Even and Paz in \cite{EvenPaz}. When there are only two agents this algorithm corresponds to ``Cut and Choose". When there are $n\geq 3$ agents, this algorithm uses a recursive strategy and it is sometimes called ``Divide and Conquer". Some properties of this approach are studied in \cite{BramsJonesKlamler}.\\
  However, it seems that one property has never been studied. Indeed, in the ``Cut and Choose" algorithm the second agent do not give his or her preference. A partition $X=X_1 \sqcup X_2$ is given and the second player choose $X_1$ or $X_2$. The measure $\mu_2$ is not used for the construction of the partition. Thus, with two agents, even if one player do not participate to the construction of the partition, we can get a proportional division. In this note, we are going to show that when there are $n+1$ agents we can construct a proportional fair division with connected portions even if the measure of one agent is unknown. We call this agent the \emph{secret agent}.\\
  
  One application suggested by the article \cite{Asada_etal} is the following: During a birthday party with $n$ guests and one host, a cake is divided into $n+1$ pieces before it is presented to the birthday boy or girl, but he or she gets any portion. 
   In this article, we give an algorithm which returns a partition assuring $1/(n+1)$ of the cake to each persons (the $n$ guests and the host) for his or her own measure.\\
   More precisely we are going to prove:
  
\begin{Thm}\label{thm:main}
Consider $n+1$ players and $X=[0,1]$.\\
We denote by $\mathcal{E}_j$ the set $\{1,2, \ldots, n, n+1\} \setminus \{j\}$.\\ 
In the Roberston-Webb model, there exists an algorithm using only queries with the first $n$ players and  giving a fair division $X=\sqcup_{j=1}^{n+1} X_j$ such that:
\begin{itemize}
\item each $X_j$ is an interval,
\item for all $j \in \{1,2, \ldots, n, n+1\}$ there exists a bijection $\sigma_j$  from $\{1,\ldots,n\}$ to $\mathcal{E}_j$  such that for all $ i \in \{1, \ldots,n\}$
$$\mu_i\big(X_{\sigma_j(i)}\big) \geq \dfrac{\mu_i(X)}{n+1}.$$
\end{itemize}
Furthermore, this algorithm uses at most $\bigO\big(n\log(n)\big)$ queries.
\end{Thm}
 
 This theorem says that the first $n$ players can construct a partition $X=\sqcup_{j=1}^{n+1} X_j$ without asking any queries to the $n+1$-th agent. Then, this last agent (the secret agent) choose first a portion $X_j$. Therefore, this agent can choose a portion $X_j$ such that  $\mu_{n+1}(X_j) \geq \mu_{n+1}(X_i)$ for $i \neq j$ and then $\mu_{n+1}(X_j) \geq 1/(n+1)$. The second part of the theorem says that the remaining portions $X_i$ can be allocated to the other agents in such a way that each agent obtained at least $1/(n+1)$ for his or her own measure.\\
 This theorem asserts that we can obtain a proportional fair division in the absence of full information. Indeed, no query is asked to the $n+1$-th player. The preferences of this player are secret. We do not know at the end of the division what is the value of $\mu_{n+1}(X_j)$. This is the reason why we call this last agent a secret agent.\\

 The ``Divide and Conquer" algorithm presented by Even and Paz in \cite{EvenPaz} \emph{ask queries to all agents} and then cannot be used to prove Theorem~\ref{thm:main}.\\
  Indeed, if there are $4$ agents this algorithm ask to \emph{each} agent the query $cut_i(0,1/2)$. This gives four cut-points $c_i$. We can suppose without loss of generality that $c_1 \leq c_2 \leq c_3 \leq c_4$. Then the first and the second agent apply the ``Cut and Choose" algorithm on $[0,c_2]$ and the third and fourth agent do the same on $[c_2,1]$. The first step of this algorithm prevent its use with a secret agent. However, we will see in the next section that this approach can be modified and used with a secret agent.\\
Other protocols can be modified and used with a secret agent. However, the complexity of these algorithms will be bigger than $\bigO\big( n \log(n)\big)$. The benefit of the modification of the ``Divide and Conquer" protocol is the following: it leads to an optimal algorithm. Indeed, it has been proved in \cite{Woeg,EdmondsPruhs} that a proportional fair division needs at least $\bigO\big(n \log(n)\big)$ queries in the Robertson-Webb model.\\

We remark that our proportional fair division method cannot be generalized with more than one secret agent. Indeed, suppose that we have $n+2$ agents and $2$ agents are secret agents, i.e. we cannot ask queries to them, they do not participate to the construction of the division. Suppose also that the $n$ agents give a partition $X=\sqcup_{j=1}^{n+2}X_j$. If the two secret agents have measure $\mu_{n+1}$, $\mu_{n+2}$ such that $\mu_{n+1}(X_1)=\mu_{n+2}(X_1)=1$, then it is impossible to obtain a proportional fair division for these $n+2$ agents with this partition.\\

We can also remark that we cannot obtain the same result for envy-free division. Indeed, Stromquist in \cite{Stromquist} has proved that there exist no algorithm in the Roberston-Web model giving a simple (connected) and envy-free fair division for $n\geq 3$ players.\\
 However, it has been proved in \cite{Asada_etal} that simple and  envy-free divisions theoretically exist even if the preferences of one person are secret. This means that this kind of partition exists but cannot be obtained with an algorithm in the Robertson-Webb model.\\

In the next section,  we give a slightly modified version of the ``Divide and Conquer" algorithm in order to prove Theorem~\ref{thm:main}.\\

\section{A modified ``Divide and Conquer" algorithm}\label{sec:proof}

\textsf{DC secret}\\
\textsf{Inputs:} $X=[a,b]$, a list $l=[\mu_1, \ldots, \mu_n]$ of $n$ players.\\
\textsf{Outputs:} A partition of $X$.\\

\begin{enumerate}
\item If $n=1$ then $c_1:=cut_1(a,1/2)$.\\
Return($X= [a,c_1] \cup [c_1,b]$).\\

\item If $n>1$ is odd then\\
\hspace{1cm} For $i$ from $1$ to $n$ do\\
\hspace{1cm} $c_i:=cut_i(a,\mu_i(X)/2)$\\
\hspace{1cm} End For.\\
Sort the set $\{c_1,\ldots,c_n\}$ in order to have: $c_{i_1} \leq c_{i_2}\leq \cdots \leq c_{i_n}.$\\
Set $X_L:=[a,c_{\frac{n+1}{2}}]$, $X_R=[c_{\frac{n+1}{2}},b]$, $l_1:=[\mu_{i_1}, \ldots, \mu_{i_{\frac{n+1}{2}-1}}]$, $l_2:=[\mu_{i_{\frac{n+1}{2}+1}},\ldots, \mu_{i_n}]$.\\
Return$\big($\textsf{DC secret}($X_L,l_1$)$\sqcup $ \textsf{DC secret}($X_R,l_2$)$\big)$.\\

\item If $n>1$ is even then \\
\hspace{1cm} For $i$ from $1$ to $n$ do\\
\hspace{1cm} $c_i:=cut_i\Big(a,\frac{\mu_i(X)}{n+1}\Big)$,\\
\hspace{1cm} End For.\\
Compute $c_{i_0}:=\min_{i=1, \ldots,n}(c_i)$.\\
Set $X_L:=[a,c_{i_0}]$ and $l':=l \setminus [\mu_{i_0}]$.\\
Return$\big(X_L \sqcup $ \textsf{DC secret}($[c_{i_0},b],l'$)$\big)$.

\end{enumerate}

\begin{proof}
We are going to prove by induction Theorem \ref{thm:main}.
We consider the following claim:\\
$H_n$: The algorithm \textsf{DC secret} applied with $n$ measures returns a partition satisfying the conclusion of Theorem \ref{thm:main}.\\

For $n=1$, $H_1$ is true. Indeed, in this case the algorithm \textsf{DC secret} gives the same partition as the ``Cut and Choose" algorithm.\\

Now, we suppose that $H_{k}$ is true for $k\leq n-1$.\\

If $n$ is even then the algorithm returns $X=X_L \sqcup \textrm{\textsf{DC secret}}(X',l')$, where $X'= X \setminus X_L$ and $l'$ is a list of $n-1$ measures. By construction $X_L$ is an interval. Without loss of generality we can suppose that $l'=[\mu_2, \ldots,\mu_n]$.\\
As $H_{n-1}$ is true we get: $X=X_L \sqcup_{j=2}^{n+1}X_j$, where $X_j$ are intervals and 
 for all $j \in \{2, \ldots, n, n+1\}$ there exists a bijection $\sigma_j$ from $ \{2,\ldots, n\}$ to   $\{2,\ldots, n+1\}\setminus \{j\}$ such that for all $i$ in $ \{2,\ldots, n\}$ we have
$$\mu_i\big(X_{\sigma_j(i)}\big) \geq \dfrac{\mu_i(X')}{n}.$$
Furthermore, for $i\geq 2$, we have  $\mu_i(X_L) \leq \frac{\mu_i(X)}{n+1}$ and $\mu_i(X')=\mu_i(X)- \mu_i(X_L)$. Thus 
$$\mu_i(X') \geq \dfrac{n}{n+1}\mu_i(X).$$
It follows 
$$\mu_i( X_{\sigma_j(i)} ) \geq \dfrac{\mu_i(X)}{n+1}.$$
Moreover, by construction, we have $\mu_1(X_L) =\mu_1(X)/(n+1)$.\\

This proves the theorem when the secret agent  chooses a portion $X_j$ and $j \geq 2$.\\

It remains thus to prove the theorem when the secret agent  chooses the portion $X_L$. However, in this situation the agent with associated measure $\mu_1$ plays the role of the secret agent in \textsf{DC secret}$(X\setminus X_L,l')$ and we get the desired result.\\

If $n>1$ is odd then the algorithm returns $X= \textrm{\textsf{DC secret}} (X_L,l_1)\sqcup \textrm{\textsf{DC secret}}(X_R,l_2)$. Without loss of generality we can suppose that $l_1=[\mu_1, \ldots, \mu_{\frac{n+1}{2}-1}]$ and $l_2= [\mu_{\frac{n+1}{2}+1},\ldots, \mu_n]$. \\
By our induction hypothesis we get: 
$$\displaystyle X= \bigsqcup_{j=1}^{\frac{n+1}{2}} X_j  \bigsqcup_{j=\frac{n+1}{2}+1}^{n+1} X_j.$$
We set $X_L=\sqcup_{j=1}^{\frac{n+1}{2}} X_j$ and $X_R=\sqcup_{j=\frac{n+1}{2}+1}^{n+1} X_j$.\\
Suppose that the secret agent  chooses a portion $X_j$ where $j\leq \dfrac{n+1}{2}$, (the other situation is similar).\\
As $H_{\frac{n+1}{2}}$ is true, there exists a bijection $\sigma_1$ from the set $\{1,\ldots,\frac{n+1}{2}-1 \} $  to $\{1,\ldots,\frac{n+1}{2} \} \setminus\{j\}$ such that for all $i$ in $\{1,\ldots,\frac{n+1}{2}-1 \}$ we have 
 $$\mu_i(X_{\sigma_1(i)}) \geq \dfrac{\mu_i(X_L)}{\frac{n+1}{2}}.$$
By construction, we have for all $i \in \{1, \ldots, \frac{n+1}{2}-1\}$,  $\mu_i(X_L) \geq \mu_i(X)/2$. This gives, 
$$\mu_i(X_{\sigma_1(i)}) \geq \dfrac{\mu_i(X)}{n+1},\textrm{ for all } i \in \Big\{1, \ldots, \frac{n+1}{2}-1\Big\}.$$\\
 
In the same way, we consider the sub-cake $X_R$, players with associated measures $\mu_{\frac{n+1}{2}+1}$, \ldots, $\mu_{n}$ and the agent with measure $\mu_{\frac{n+1}{2}}$ as a secret agent. 
By construction, this last agent  is such that 
$$\mu_{\frac{n+1}{2}}(X_R)=\mu_{\frac{n+1}{2}}(X)/2.$$
 Furthermore, $X_R$ is divided in $(n+1)/2$ parts, thus there exists a portion $X_k$ such that 
 $$\mu_{\frac{n+1}{2}}(X_k) \geq \dfrac{\mu_{\frac{n+1}{2}}(X_R)}{\frac{n+1}{2}}.$$
 We deduce 
 $$\mu_{\frac{n+1}{2}}(X_k) \geq \dfrac{\mu_{\frac{n+1}{2}}(X)}{n+1}.$$

As before, as $H_{\frac{n+1}{2}}$ is true, we get:  there exists a bijection $\sigma_2$ from the set $\{ \frac{n+1}{2}+1,\ldots,n\}$ to $\{ \frac{n+1}{2}+1,\ldots,n+1\} \setminus \{k\}$, such that for $i \in \{ \frac{n+1}{2}+1,\ldots,n\}$ we have 
$$\mu_i(X_{\sigma_2(i)}) \geq \dfrac{\mu_i(X)}{n+1}.$$
From the bijection $\sigma_1$ and $\sigma_2$ we can construct a bijection $\sigma$ giving the conclusion of the theorem.\\

The complexity study is classical and is the same as the one done for the usual ``Divide and Conquer" algorithm: the number of steps is in $\bigO\big(\log(n)\big)$ and the number of queries in each step is in $\bigO(n)$. This gives the desired complexity.
\end{proof}


 \bibliographystyle{plain} 
 
\bibliography{cakebiblio}

\end{document}